\documentstyle[prb,twocolumn,aps]{revtex}
\input epsf
\begin{document}
\draft
\title{Phononic crystals with planar defects}
\author{I.~E.~Psarobas\cite{ntua} and N.~Stefanou}
\address{Section of Solid State Physics, University of Athens,
Panepistimioupolis,\\ GR-157 84, Athens, Greece}
\author{A.~Modinos}
\address{Department of Physics, National Technical University of Athens,
Zografou Campus,\\ GR-157 73, Athens, Greece}
\date{\today}
\maketitle
\begin{abstract}
\footnotesize{\bf We study the effect of planar defects in phononic
crystals of spherical scatterers. It is shown that a plane of
impurity spheres introduces modes of vibration of the elastic field
localized on this plane at frequencies within a frequency gap of a
pure phononic crystal; these show up as sharp resonances in the
transmittance of elastic waves incident on a slab of the crystal. A
periodic arrangement of impurity planes along a given direction
creates narrow impurity bands with a width which depends on the
position of these bands within the frequency gap of the pure crystal
and on the separation between the impurity planes. We show how a
slight deviation from periodicity (one impurity plane is different
from the rest) reduces dramatically the transmittance of elastic
waves incident on a slab of the crystal.}
\end{abstract}
%%%%%%%%%%%%%%%%%%%%%%%%%%%%% SECTION I %%%%%%%%%%%%%%%%%%%%%%%%%%%%%%%
\section{Introduction}
\label{intro} Phononic crystals are composite materials whose
elastic properties, as described by the mass density $\rho$ and the
Lam\'e coefficients $\lambda$ and $\mu$, vary periodically in space.
In recent years there has been a growing interest in the study of
phononic crystals, especially in relation to the possibility of
phononic band gap materials (see, e.g., Ref.~\onlinecite{paper1} and
references therein). By definition these exhibit regions of
frequency (phononic gaps) over which no vibrations are possible
within the infinite crystal. This implies that an elastic wave,
whatever its direction of propagation and its polarization
(longitudinal or transverse), incident on a slab of the material of
some thickness will be totally reflected by it, if its frequency
lies within a phononic gap. Obvious technological applications of
the above are non-absorbing mirrors and vibration-free cavities
which might be very useful in high-precision mechanical systems
operating in a given frequency range.

On the theory side one would like to be able to calculate the
frequency band structure of a phononic crystal and also the
transmission and reflection coefficients of elastic waves incident
on a slab of the material of finite thickness. In a previous
publication \cite{paper1} (to be referred to as paper I) we
presented a formalism which can do that for phononic crystals
consisting of non-overlapping spheres embedded in an elastic medium
of different mass density and Lam\'e coefficients. In the present
paper we apply the formalism of paper I to a specific system
consisting of non-overlapping solid spheres embedded in a solid host
medium. We assume that the spheres are centered on the sites of an
fcc lattice and view the crystal as a sequence of (001) planes along
the $z$ axis. We calculate the frequency band structure of the
infinite crystal, and the transmission coefficient for an elastic
wave incident on a slab of the material parallel to the (001)
surface consisting of a finite number of planes of spheres. For a
thick slab the transmittance vanishes over the exact range of the
band gap (of the infinite crystal) as one would expect; but of
greater interest is the fact that the transmittance behaves in
almost the same manner for a thin slab consisting of just two planes
of spheres.

Linear and point defects in phononic crystals have been investigated
by Torres {\em et al.}.~\cite{Torres} In the present work we study
the effect of planar defects, i.e. of impurity planes, on the
transmittance of a slab. An impurity plane has the same
two-dimensional (2D) periodicity as the rest of the planes of the
slab, but the spheres centered on this plane are different. In our
example they have a different radius. This leads to gap states:
vibrational modes of the elastic field localized on the impurity
plane, which show up as transmission resonances at frequencies
within the gap of the pure crystal.

Finally we consider a crystal whose unit cell along the $z$
direction consists of a number of planes one of which (impurity
plane) is different from the other planes of the unit cell. This
leads to impurity bands within the frequency gap of the pure
crystal, the width of which depends on the position of the impurity
band within the gap and on the separation between the impurity
planes. We show that a slight deviation from periodicity reduces
dramatically the transmission coefficient of an elastic wave
incident on a slab of the crystal.

In all calculations of the transmittance of a slab we assume that
the material that exists between the spheres of the slab extends to
infinity on either side of the slab.
%%%%%%%%%%%%%%%%%%%%%%%%%%%%% SECTION II %%%%%%%%%%%%%%%%%%%%%%%%%%%%%%%
\section{Properties of the pure crystal}
\label{properties} The pure crystal consists of non-overlapping lead
spheres centered on the sites of an fcc lattice. The elastic
properties of the spheres are characterized by the mass density,
$\rho$, and the longitudinal and transverse velocities of
propagation, denoted by $c_l$ and $c_t$ respectively. The latter are
given by: $c_l^2=(\lambda + 2\mu)/ \rho$ and $c_t^2=\mu/ \rho$. For
lead we have: $\rho=11.357 \: \mathrm{gr\;cm^{-3}}$, $c_l=2158 \:
\mathrm{m\; s^{-1}}$, $c_t=860  \: \mathrm{m\;s^{-1}}$. The space
between the spheres is filled with epoxy ( $\rho=1.180 \:
\mathrm{gr\;cm^{-3}}$, $c_l=2540 \: \mathrm{m\;s^{-1}}$, $c_t=1160
\: \mathrm{m\;s^{-1}}$).

\begin{figure}
\centering
 \epsfxsize=8cm \epsfbox{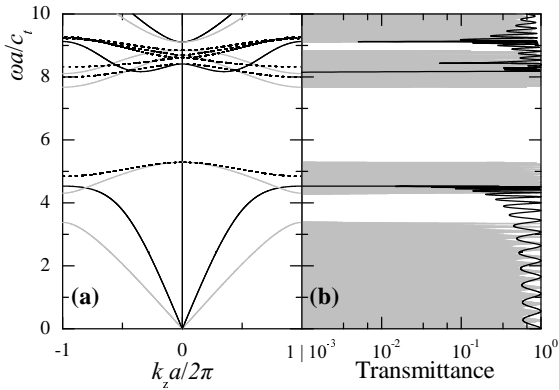}
\caption{The phononic band structure at the center (${\bf
k}_{\|}={\bf 0}$) of the SBZ of a (001) surface of an fcc crystal of
lead spheres in epoxy (a); and the corresponding transmittance curve
of a slab of 16 planes of spheres parallel to the same surface (b).
The fractional volume occupied by the spheres is $0.262$. In (a) the
black lines represent longitudinal modes, the grey lines transverse
modes, and the dotted lines are deaf bands. Correspondingly, in (b)
the black line shows the transmittance of longitudinal incident
elastic waves, and the shaded curve that of transverse incident
waves. } \label{fig1}
\end{figure}

We calculate the complex frequency band structure of the infinite
crystal (extending from $z \rightarrow - \infty$ to $z \rightarrow
\infty$) associated with the (001) surface in the manner described
in paper I. For any given reduced wavevector ${\bf k}_{\parallel}$,
within the surface Brillouin zone (SBZ) of this surface, we
calculate the frequency lines $k_z(\omega;{\bf k}_{\parallel})$ as
functions of the angular frequency $\omega$. Our results, obtained
with an angular momentum cut-off $\ell_{max}=4$ and 13 2D reciprocal
vectors, ${\bf g}$, are converged within an accuracy $10^{-3}$. The
ordinary frequency band structure which interests us here,
corresponds to real sections (regions of frequency over which
$k_z(\omega;{\bf k}_{\parallel})$ is real) of these lines. In
Fig.~\ref{fig1}(a) we show the frequency bands of the elastic field
for ${\bf k}_{\parallel}={\bf 0}$ (along the direction normal to the
(001) plane). We present our results in dimensionless units; $a$
denotes the lattice constant of the fcc lattice and $c_t$ the
propagation velocity of transverse elastic waves in epoxy. The
radius $S$ of the spheres equals $0.25a$ in  the present case which
corresponds to a fractional volume occupied by the spheres, denoted
by $f$, equal to 0.262.
%%%%%%%%%%%%%%%%%%%FIGURE 1%%%%%%%%%%%%%%%%%%%%%%%%%%%%%%%%%%%%%%%%%
The bands shown in Fig.~\ref{fig1}(a) by black (grey) lines
correspond to longitudinal (transverse) modes of vibration of the
elastic field inside a sufficiently thick slab of the material.
These modes are excited only by longitudinal (transverse) waves
incident normally on the surface of the slab. This is demonstrated
quite clearly in Fig.~\ref{fig1}(b), which shows the transmission
coefficient of a plane wave incident normally on a slab of the
material consisting of 16 planes of spheres parallel to the (001)
surface. The shaded curve shows the transmission coefficient for a
transverse incident wave and the black line that of a longitudinal
incident wave.
%%%%%%%%%%%%%%%%%%%%FIGURE 2%%%%%%%%%%%%%%%%%%%%%%%%%%%%%%%%%%%%%
\begin{figure}
\centering
 \epsfysize=8cm \epsfbox{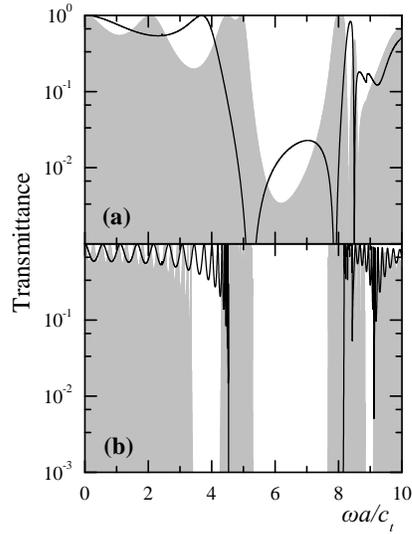}
\caption{ Transmittance of longitudinal (black lines) and transverse
(shaded curves) elastic waves incident normally on a slab of the
phononic crystal described in the caption of Fig.~\ref{fig1},
consisting of two planes of spheres (a), and 16 planes of spheres
(b).} \label{fig2}
\end{figure}
We hasten to add (see also paper I) that in a composite material the
modes of vibration of the elastic field need not be purely
longitudinal or purely transverse even along symmetry directions
and, therefore, the fact that they behave so in the system under
consideration is worth noting. Off the normal direction (${\bf
k}_{\parallel} \neq {\bf 0}$) the frequency bands are clearly
hybridized and they are excited by either longitudinal or transverse
incident plane waves. The bands shown by dotted lines in
Fig.~\ref{fig1}(a) are deaf bands. The ${\bf g}={\bf 0}$ component
of the corresponding modes of vibration vanishes and that being the
only component which couples with the external elastic field in the
frequency range under consideration, they cannot be excited by an
incident plane wave, leading to total reflection of the latter.
Finally, we point out the oscillations in the transmission
coefficient, over the allowed regions of frequency, which are due to
interference effects resulting from multiple reflection between the
surfaces of the slab (Fabry-Perrot-type oscillations).

In Fig.~\ref{fig2} we show the transmission coefficient of an
elastic wave incident normally on a thin (001) slab consisting of
just two planes of spheres, and compare it with the same quantity
for a thick slab (of 16 planes of spheres), which as we have seen is
determined by the frequency band structure of the infinite crystal.
We see that over the frequency region that constitutes the frequency
gap of the infinite crystal, the transmittance of a slab consisting
of just two planes of spheres is not zero, but it is nevertheless
about two orders of magnitude smaller than its value at frequencies
below or above the gap edges. The Fabry-Perrot-type oscillations are
of course different in the two cases since they depend directly on
the thickness of the slab.
%%%%%%%%%%%%%%%%%%%%FIGURE 3%%%%%%%%%%%%%%%%%%%%%%%%%%%%%%%%%%%%%%
\begin{figure}
\centering
 \epsfxsize=8cm \epsfbox{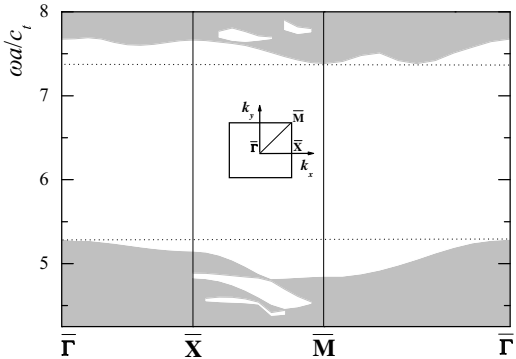}
\caption{ Projection of the frequency band structure on the SBZ of
the (001) surface of the fcc phononic crystal described in the
caption of Fig.~\ref{fig1}. The blank areas show the frequency gaps
in the considered frequency region. The inset shows the SBZ of the
(001) fcc surface. } \label{fig3}
\end{figure}

An absolute frequency gap extends between $\omega_l a/c_t = 5.29$
and $\omega_u a/c_t = 7.38$, as demonstrated in Fig.~\ref{fig3}.
This shows the projection of the frequency band structure on the
SBZ, along the symmetry lines of the latter shown in the inset. The
shaded areas correspond to frequencies for which there exists at
least one propagating Bloch wave (an eigenmode of the elastic field)
in the infinite crystal; in the frequency regions represented by the
blank areas none exists. We verified, by calculating the frequency
lines at selected points of the SBZ, that the gap does exist over
the entire SBZ, and that it is therefore an absolute phononic gap
extending from $\omega_l a/c_t = 5.29$ to $\omega_u a/c_t = 7.38$.
The above values of the lower and upper edges of the gap are
different from the values $\omega_l a/c_t \approx 6.2$ and $\omega_u
a/c_t \approx 8.0$, respectively, found by Kafesaki {\it et
al.},~\cite{kaf95} who studied the same system using the plane-wave
method. We believe our results to be the more accurate. It is now
recognized that the plane-wave method is not particularly suited to
the study of phononic crystals consisting of non-overlapping spheres
in a host medium, because of the very large number of plane waves
required to obtain convergent results.~\cite{kaf99}
%%%%%%%%%%%%%%%%%%%%%%%%%%%%% SECTION III %%%%%%%%%%%%%%%%%%%%%%%%%%%%%%%
\section{Planar defects}
\label{defects} 
\begin{figure}
\centering
 \epsfxsize=7cm \epsfbox{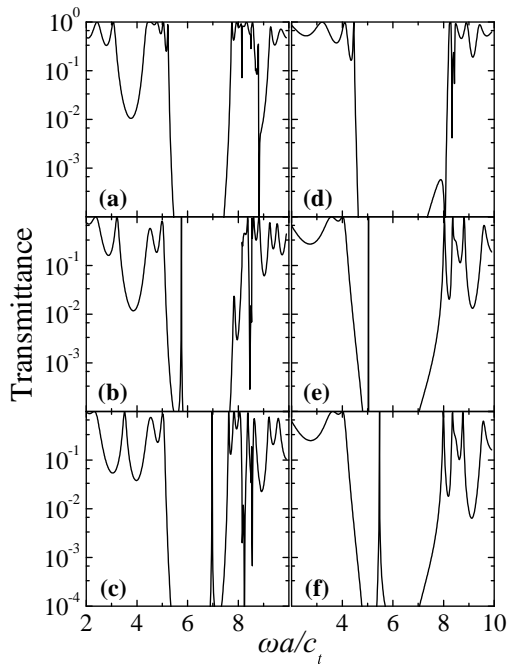}
\caption{ Transmittance of elastic waves incident normally on a slab
of an fcc crystal of lead spheres in epoxy. The slab consists of
five planes of spheres parallel to the (001) surface. The spheres
have a radius $S=0.25a$, except those of the middle plane which have
a different radius $S_i$ [(a): $S_i=S$, (b): $S_i=0.8 S$, (c):
$S_i=0.6 S$, for a transverse incident wave; (d): $S_i=S$, (e):
$S_i=0.6 S$, (f): $S_i=0.5 S$, for a longitudinal incident wave].}
\label{fig4}
\end{figure}

Next we consider a slab of the phononic crystal
consisting of five planes of spheres, one of which may be different
from the other four. It has the same 2D periodicity parallel to the
surface of the slab, but the spheres of this, so-called, impurity
plane may be different: they have a smaller or larger radius than
the spheres of the other planes. Our results are summarized in the
six diagrams of Fig.~\ref{fig4}. The first three, (a), (b) and (c),
show the transmittance of a transverse elastic wave incident
normally on a slab of five planes of spheres. For Fig.~\ref{fig4}(a)
all five planes are the same, as in the infinite crystal described
in the preceding paragraphs, and the transmittance vanishes, as
expected, over the frequency region corresponding to the frequency
gap of the infinite crystal for ${\bf k}_\| = {\bf 0}$, shown in
Fig.~\ref{fig1}(a). For Figs.~\ref{fig4}(b) and \ref{fig4}(c) the
middle plane is different from the other four. The spheres of this
plane have a smaller radius: $S_i=0.8S$ for Fig.~\ref{fig4}(b) and
$S_i=0.6S$ for Fig.~\ref{fig4}(c), where S is the radius of the
spheres in the other planes. The transmission resonance that now
appears at a frequency within the gap signifies the existence of a
state of the elastic field centered on the impurity plane: a mode of
vibration of the elastic field that extends to infinity parallel to
the surface of the slab (in the manner of a Bloch wave), but decays
rapidly normal to the impurity plane on either side of it. It
appears that the normal mode of vibration (with ${\bf k}_{\parallel}
= {\bf 0}$) at the top of the valence band (we use the term by
analogy to semiconductor physics to denote the frequency band below
the gap), splits off this band, becoming a localized (on the
impurity plane) vibration, with a frequency higher in the gap the
smaller the radius of the impurity spheres. Figs. \ref{fig4}(d) to
\ref{fig4}(f) demonstrate the same phenomenon for longitudinal waves
incident normally on the slab. For Fig.~\ref{fig4}(d) all five
planes are the same as in the infinite crystal, and the transmission
coefficient accords with the corresponding frequency band structure
shown in Fig.~\ref{fig1}(a). For Figs.~\ref{fig4}(e) and
\ref{fig4}(f) the spheres of the middle plane have a radius smaller
than that of the other four planes: $S_i=0.6S$ for
Fig.~\ref{fig4}(e) and $S_i=0.5S$ for Fig.~\ref{fig4}(f). We note
that while the trend is the same, the resonance frequency stays
closer to the top of the valence band for longitudinal waves.

\begin{figure}
\centering
 \epsfxsize=7cm \epsfbox{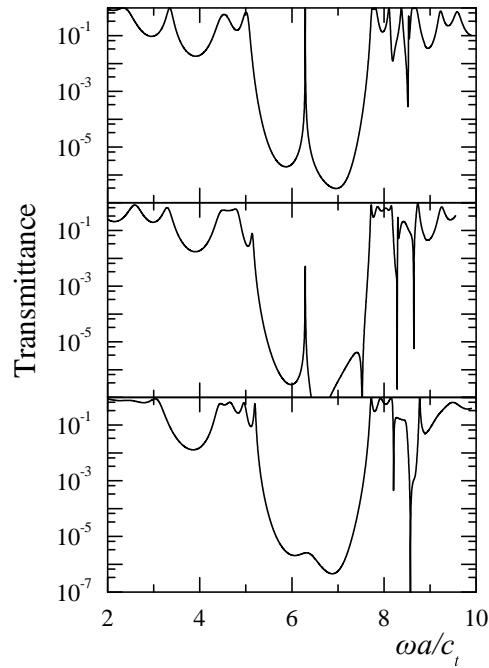}
\caption{Transmittance of transverse elastic waves incident normally
on a slab of an fcc crystal of lead spheres in epoxy. The slab
consists of five planes of spheres parallel to the (001) surface.
The spheres have a radius $S=0.25a$, except those of the middle
plane (top diagram), or of the second plane from the surface (middle
diagram), or of the surface of the slab (bottom diagram), which have
$S_i=0.7S$ .} \label{fig5}
\end{figure}

\begin{figure}
\centering
 \epsfxsize=8cm \epsfbox{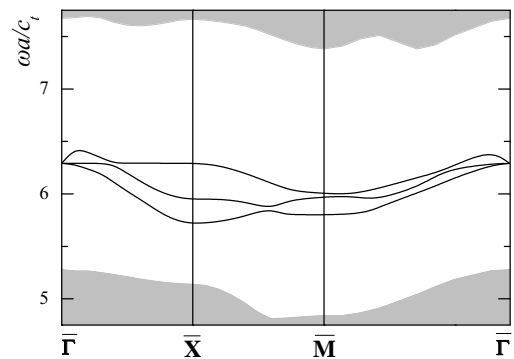}
\caption{Variation with ${\bf k}_{\|}$ of the resonance frequency
associated with an impurity plane at the center of a slab of five
(001) fcc planes of lead spheres in epoxy ($S=0.25a$, $S_i=0.7S$).}
\label{fig6}
\end{figure}

We note that the transmission coefficient equals unity at the
resonance frequency. This is the case when the impurity plane lies in
the middle of the slab; when the impurity plane is any other than the
middle plane, the transmission coefficient at the resonance frequency
is less than unity, as demonstrated in Fig.~\ref{fig5}. This figure
shows the transmission coefficient for a transverse plane wave
incident normally on a slab of five planes of spheres one of which is
an impurity plane. The spheres centered on this plane have a radius
$S_i=0.7S$, where $S$ is the radius of the spheres of the other
planes. The top diagram is obtained when the impurity plane is the
middle plane of the slab, the middle diagram is obtained when the
impurity plane is the second from the surface, and the third diagram
is obtained when the impurity plane lies at the surface of the slab.
Only in the first case is the transmission coefficient at resonance
equal to unity. When the impurity plane is removed from the center of
the slab by one plane, the value of the transmission coefficient at
resonance diminishes by at least two orders of magnitude (middle
diagram), and the resonance disappears altogether when the impurity
plane is removed to the surface of the slab. Similar results are
obtained for a longitudinal incident wave. This is indeed a general
characteristic of resonant tunneling. We note, for example, the
similarity of the above results with those obtained in the
transmission of an electron through a double barrier, when the well
between the two barrier tops allows for a resonance state of the
tunneling electron. When the double barrier is symmetric the
transmission coefficient at resonance equals unity, when the double
barrier is not symmetric the transmission coefficient is less than
unity.~\cite{mod69}
In Fig.~\ref{fig6} we show the variation with ${\bf k}_{\parallel}$
of the resonance frequency, associated with an impurity plane at the
middle of a slab of five planes of spheres. The spheres of the middle
plane have a radius $S_i = 0.7S$, where $S=0.25a$ is the radius of
the spheres in the other planes. The gap states/resonances along
$\bar{\rm X} \bar{\rm M}$ and $\bar{\rm M} \bar{\rm \Gamma}$ are
hybridized, excited by both longitudinal and transverse incident
plane waves. The states/resonances along $\bar{\rm \Gamma} \bar{\rm
X}$ corresponding to the two top bands are excited by $s$-polarized
transverse waves ( the displacement vector is parallel to the surface
of the slab); the third band is excited by both longitudinal and
$p$-polarized waves (the displacement vector lies in the plane of
incidence). We note that all three bands of gap states are restricted
within a rather narrow range of frequencies about the midgap
frequency.
%%%%%%%%%%%%%%%%%%%%FIGURE 5%%%%%%%%%%%%%%%%%%%%%%%%%%%%%%%%%%%%%%%

%%%%%%%%%%%%%%%%%%%%FIGURE 6%%%%%%%%%%%%%%%%%%%%%%%%%%%%%%%%%%%%%%%

\begin{figure}
\centering
 \epsfxsize=8cm \epsfbox{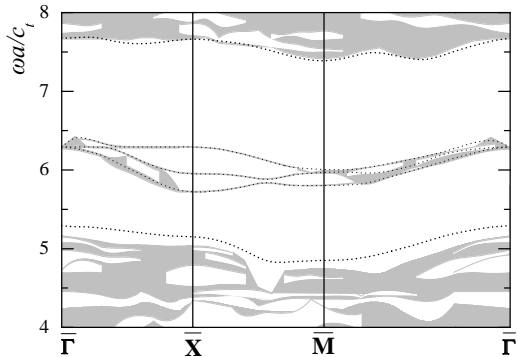}
\caption{Projection of the frequency band structure on the SBZ of
the (001) fcc surface of a phononic crystal of lead spheres in
epoxy. The spheres are centered on the sites of an fcc lattice and
the unit cell of the structure along the $z$ axis consists of three
planes of spheres: $SSS_i$, with $S=0.25a$ and $S_i=0.7S$. The
dotted lines reproduce the results shown in Fig.~\ref{fig6} for
comparison.} \label{fig7}
\end{figure}

Next we consider an infinite crystal (extending from $z \rightarrow
-\infty$ to $z \rightarrow \infty$) the unit cell of which, along
the $z$ axis, consists of three fcc planes of spheres; we have:
$\cdots SSS_iSSS_iSSS_i \cdots$. The spheres of the third plane of
the unit slice have a radius $S_i$ which is smaller than the radius
$S$ of the spheres of the other two planes. We put $S_i = 0.7S$. The
resonances shown in Fig.~\ref{fig6} now develop into bands shown by
shaded regions in Fig.~\ref{fig7}. These bands are nevertheless
extremely narrow for most ${\bf k}_{\parallel}$, which implies that
the vibrations are strongly localized on the impurity planes, with
very little interaction between impurity planes. This is, of course,
consistent with the sharpness of the resonance seen in a slab of
five planes (see top diagram of Fig.~\ref{fig5}).
%%%%%%%%%%%%%%%%%%%%FIGURE 7%%%%%%%%%%%%%%%%%%%%%%%%%%%%%%%%%%%%%%%
We can vary the width of the impurity bands by pushing them away
from the middle of the gap. Putting $S_i=0.8 \,S$ displaces the
impurity bands of Fig.~\ref{fig7} towards the bottom of the gap; at
the $\bar{\rm \Gamma}$ point this leads to a band-width of $\Delta
\omega\, a/c_t\approx 0.14$ which is more than three times that
shown in Fig.~\ref{fig7} which equals $0.04$. We can understand this
by noting that the interaction between states localized on
neighboring impurity planes depends on the separation, $d$, between
the planes approximately as $\sim \exp \left[-|\,k_{i}(\omega;{\bf
k}_{\|} )|\,d\,\right]$, where $k_i(\omega;{\bf k}_{\|})$ is the
smallest (in magnitude) of the imaginary parts of the frequency
lines $k_z(\omega;{\bf k}_{\|})$ in the pure crystal over the
frequency range of the impurity band. $|\,k_i(\omega;{\bf k}_{\|})|$
is largest in the middle of the gap (for a given ${\bf k}_{\|}$) and
goes to zero at the edges of this gap.

\begin{figure}
\centering
 \epsfxsize=9cm \epsfbox{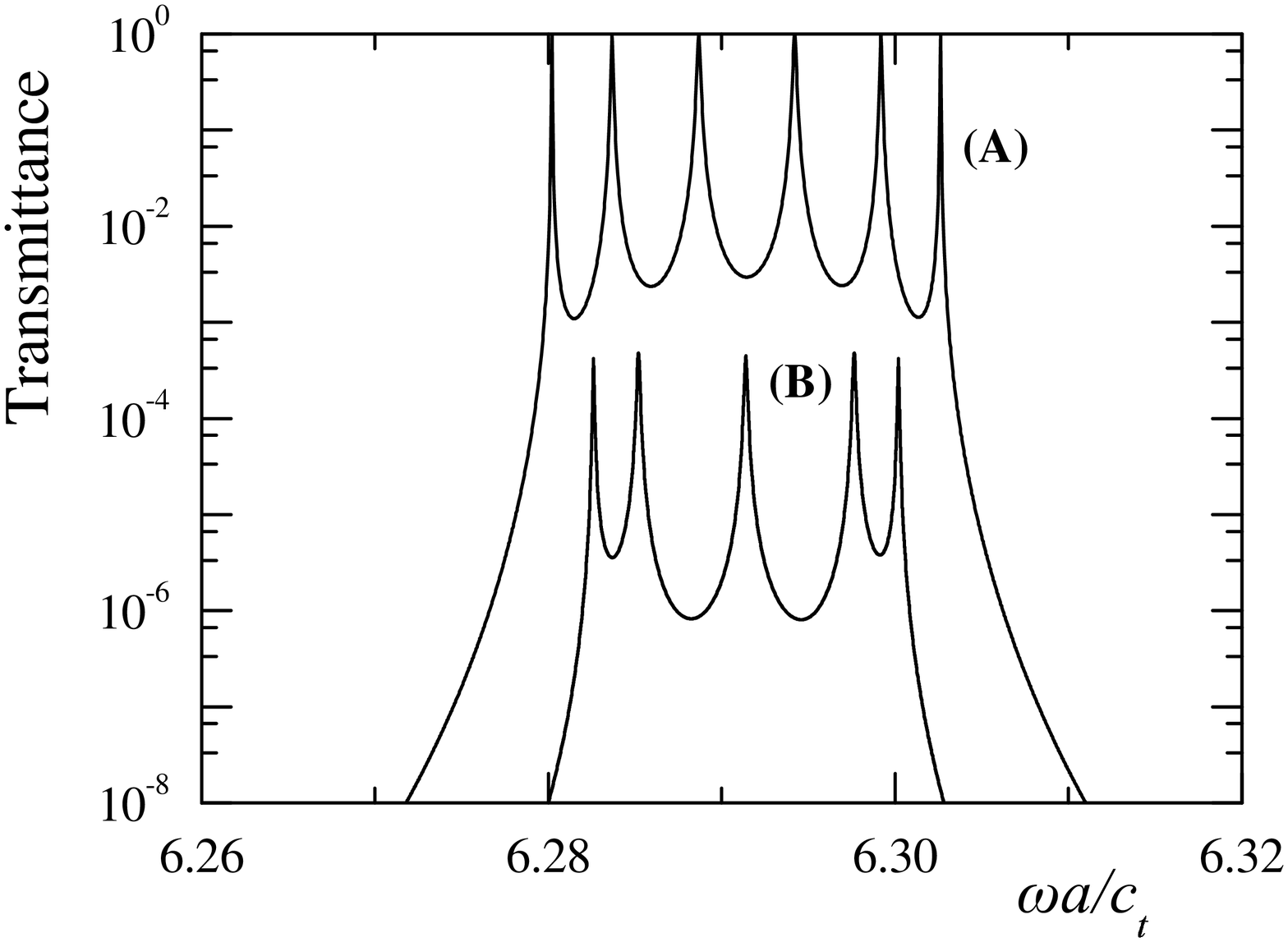}
\caption{(A): Transmittance of transverse elastic waves incident
normally on a slab of seven unit slices of the phononic crystal
described in the caption of Fig.~\ref{fig7}, over the frequency
range of the impurity band within the absolute gap. (B): As in (A)
except that $S_i$ of the third unit slice is a little different (it
equals $0.6S$).} \label{fig8}
\end{figure}

Finally, in Fig.~\ref{fig8} (curve (A)) we show the transmission
coefficient in the frequency region of the impurity band for a
transverse wave incident normally on a slab of a phononic crystal
consisting of seven unit cells along the $z$ direction, when the
unit slice consists of three (001) fcc planes: $SSS_i$, with the
spheres of the third plane having a radius $S_i=0.7S$ with
$S=0.25a$.
%%%%%%%%%%%%%%%%%%FIGURE 8%%%%%%%%%%%%%%%%%%%%%%%%%%%%%%%%%%%%%%%%%%
We note that the frequency gap of the pure crystal ($S_i=S$) for
${\bf k}_{\|}={\bf 0}$ extends from $\omega_l\ a/c_t=5.29$ to
$\omega_u a/c_t=7.67$. We recall that the gap states corresponding to
the first six impurity planes (taken individually) manifest
themselves as transmission resonances at $\omega_r a/c_t\approx6.29$
(see Fig.~\ref{fig5}) while the seventh impurity plane, at the
surface of the slab, does not give a transmission resonance, as shown
in the bottom diagram of Fig.~\ref{fig5}. The interaction between the
impurity planes removes the degeneracy of the above resonance level,
leading to six discrete resonances in the transmission spectrum, as
shown in Fig.~\ref{fig8}. Curve (B) of Fig.~\ref{fig8} is obtained
for a slab which differs from the above in that the spheres of the
third plane of the third unit slice have a radius $S_i=0.6S$, instead
of $S_i=0.7S$. We note the dramatic reduction, by about four orders
of magnitude, in the transmission coefficient as a result of the
presence of just one different impurity plane. In a way this
phenomenon, which may have interesting technological applications, is
to be expected; the different impurity plane decouples the vibrations
on its right from those on its left. This suggests that a random
distribution of impurity planes will lead to vibrational modes
localized over smaller regions of the slab. We have here a
possibility to study Anderson localization of classical waves in an
effectively one-dimensional system (because of ${\bf k}_{\|}$
conservation) which allows direct comparison with corresponding
experimental data, without the difficulty associated with correlation
in the corresponding electronic problem.\newline

%%%%%%%%%%%%%%%%%%%%%%%%%%%%%%% FIGURES %%%%%%%%%%%%%%%%%%%%%%%%%%%%
\end{document}